\documentclass[%
 reprint,
 amsmath,amssymb,
 aps, pre
]{revtex4-2}

\usepackage{graphicx}
\usepackage{dcolumn}
\usepackage{bm}
\usepackage[mathlines]{lineno}

\usepackage{xcolor}

\begin{document}

\preprint{APS/123-QED}

\title{Nonuniqueness of Tangential and Kinematic Restitution in Frictional Oblique Particle Impacts}

\author{Dominik Krengel}
 \email{dominik.krengel@kaiyodai.ac.jp}
\affiliation{
 Department of Marine Resources and Energy, \\
 Tokyo University of Marine Science and Technology,\\
 4-5-7, Konan, Minato, 108-8477, Tokyo, Japan
}

\date{\today}

\begin{abstract}
 We systematically investigate the different coefficients of restitution of a dissipative, frictional disc as a function of the impact angle $\theta$, the friction coefficient $\mu$ and initial rotation $\omega_{\mathrm{in}}$. We observe a non-monotonic, non-linear behaviour of $e_{\mathrm{T}}$  and $e_{\mathrm{kin}}$ with a clear minimum at $\mu$-dependent values of $\theta$ and a convergence for all $\mu$ at large $\theta$, bounded by the cases of pure rolling and pure sliding. Changing the dissipative normal interaction affects the $\mu$ convergence at large $\theta$ and leads to a convergence at low $\theta$. The presence of initial angular velocity $\omega_{\mathrm{in}}$ can significantly alter the functional behaviour of $e_{\mathrm{T}}$ at steep impacts and slightly change the magnitude of $e_{\mathrm{kin}}$, with friction playing only a minor role. Overall, our results indicate that any specific value of $e_{\mathrm{T}}$ or $e_{\mathrm{kin}}$ is highly situational and can simultaneously describe entirely different contact configurations.
\end{abstract}

\maketitle
\textsl{Introduction.} 
Particle impacts are a simple problem: a particle hits another body, exchanges energy and momentum, and possibly separates again. In the absence of internal degrees of freedom, the collision involves three rectilinear degrees of freedom (DOF) and three angular degrees of freedom. For spherical bodies the problem can be simplified to one normal and one tangential DOF, together with one additional DOF for rotation. Restitution coefficients $e$ are typically used to relate the pre- and post-contact kinematic variables of their associated DOF,
\begin{equation}
 \Delta \nu' = \pm e\cdot\Delta\nu,
 \label{eq:restitution}
\end{equation}
where $\Delta\nu$ is the pre- and $\Delta\nu'$ is the post-contact velocity. The sign of the right-hand side in Eq.\,\ref{eq:restitution} is negative  for the coefficient of normal restitution $e_{\mathrm{N}}$, where $\Delta\nu = \Delta v_{\mathrm{N}}$, and positive for the tangential restitution coefficient $e_{\mathrm{T}}$ , where $\Delta\nu = \Delta v_{\mathrm{T}}$ and the kinematic restitution coefficient $e_{\mathrm{kin}}$ with $\Delta v = \sqrt{\Delta v_{\mathrm{N}}^2 + \Delta v_{\mathrm{T}}}^2$\,\cite{Schwager2007, Becker2008, Asteriou2018}.

Fields like the kinetic theory of granular gases\,\cite{Brilliantov2004,Brilliantov2007}, efficient {E}vent-{Driven} numerical simulations\,\cite{Poeschel2005, Wedel2024} and rockfall modelling\,\cite{Chau2002,Tang2021} all are built upon the various definitions of $e$. The different $e$ are determined from experiment, typically less than unity due to the dissipation of energy during the interaction. Historically, they have been assumed either to be constant or monotonically decreasing with the collision velocity. However recent research has painted a more complex picture of the normal component $e_{\mathrm{N}}$. Oblique collisions between two discs may lead to $e_{\mathrm{N}} < 0$ as the particles rotate around their mutual center\,\cite{Saitoh2010,Mueller2012}. In contrast, in oblique impacts with a plate local deformation and transfer from internally stored energy can result in $e_{\mathrm{N}} > 1$\,\cite{Louge2002, Kuninaka2004}, while the velocity dependence of $e_{\mathrm{N}}$ has been shown to be non-monotonic\,\cite{Mueller2013}. 

However, the overwhelming majority of studies have focused only on normal restitution $e_{\mathrm{N}}$. Although equally important in describing collisions, the dependencies of the tangential and kinematic restitution coefficients are much less understood. Likewise, laboratory measurements are constrained by, e.g. deformation and contamination of surfaces, and difficult to control interdependence of setup parameters, limiting direct implementation in numerical simulations. Likewise, analytical formulations tend to omit the influence of particle rotation at the contact, although the inter-dependence between the different degrees of freedom is well known\,\cite{Johnson1985,Kalker1990,Farkas2003,Yu2017,Santos2020,Jenkins2026}.

The purpose of this article is to clarify the relation between $e$, the dissipative parameters in normal and tangential direction, and initial angular velocity in oblique impacts of discs on a horizontal plate by means of two-dimensional Discrete Element simulations. We provide a systematic mapping of the resulting non-uniqueness in $e_{\mathrm{T}}$ and $e_{\mathrm{kin}}$ across impact angle, friction coefficient and initial rotation.

\begin{figure}[b]
 \centering
 \includegraphics[width=0.8\columnwidth]{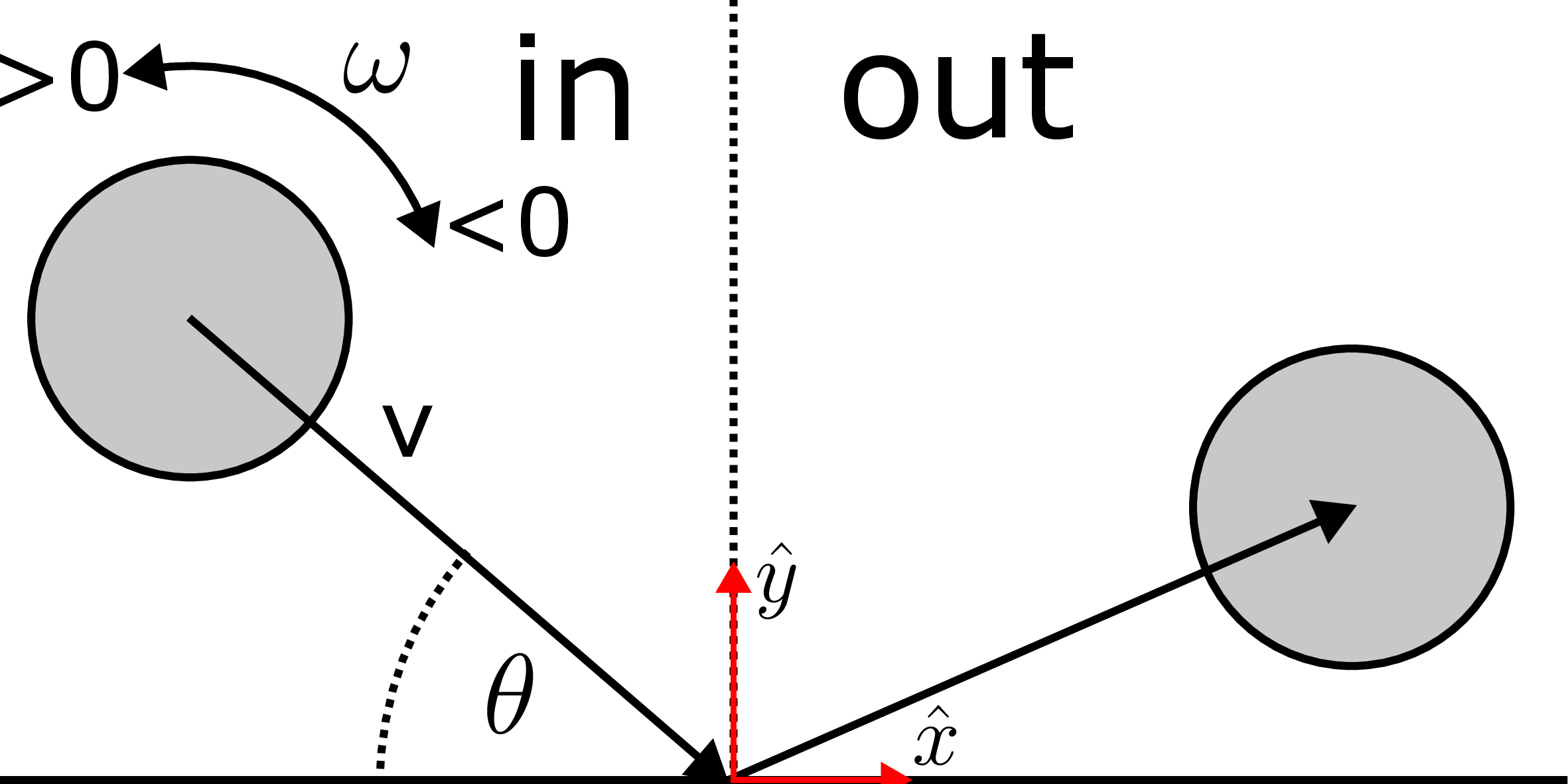}
 \caption{Simulation configuration.}
 \label{fig:01_setup}
\end{figure}
\textsl{Simulation.} We perform {D}iscrete {Element} simulations (DEM) of the planar impact of a disc with radius $r$ under varying impact angles $\theta$ and friction coefficients $\mu$ (Fig.\,\ref{fig:01_setup}). The initial rectilinear velocity magnitude $v$ is constant for all cases so that $\left(v_x(0),v_y(0)\right)^{\top} = v\left(\cos{\theta}, \sin{\theta}\right)^{\top}$, with the impact angle $4^{\circ}\leq \theta\leq89^{\circ}$. This means, in contrast to typical experiments, only the ratio between impact normal and tangential velocity is fixed, not their individual values. The initial angular velocity $\omega=0$, unless specified otherwise. Upon contact friction (proportional to a coefficient $\mu$) acts in tangential direction ($\parallel \hat{x}$), dissipation (proportional to a coefficient $\gamma$) acts in normal direction ($\parallel \hat{y}$). Both the particle and the plate have a finite Young's modulus, so the particle numerically penetrates the plate during contact as a stand-in for deformation, but no internal degrees of freedom (e.g. vibration) are allowed. Gravity is disabled during the simulation to remove its effect on impact trajectory and velocity. The induced forces are proportional to the area of the overlap polygon. The particle and the plate are treated as rigid, no internal degrees of freedom or plastic deformation are being considered. The simulation algorithm is described in more detail in the supplementary material, the setup in\,\cite{Krengel2026}. We compute the restitution coefficients based on the velocities measured after the particle separates from the plane.

\begin{figure}[b!]
 \centering
 \includegraphics[width=\columnwidth]{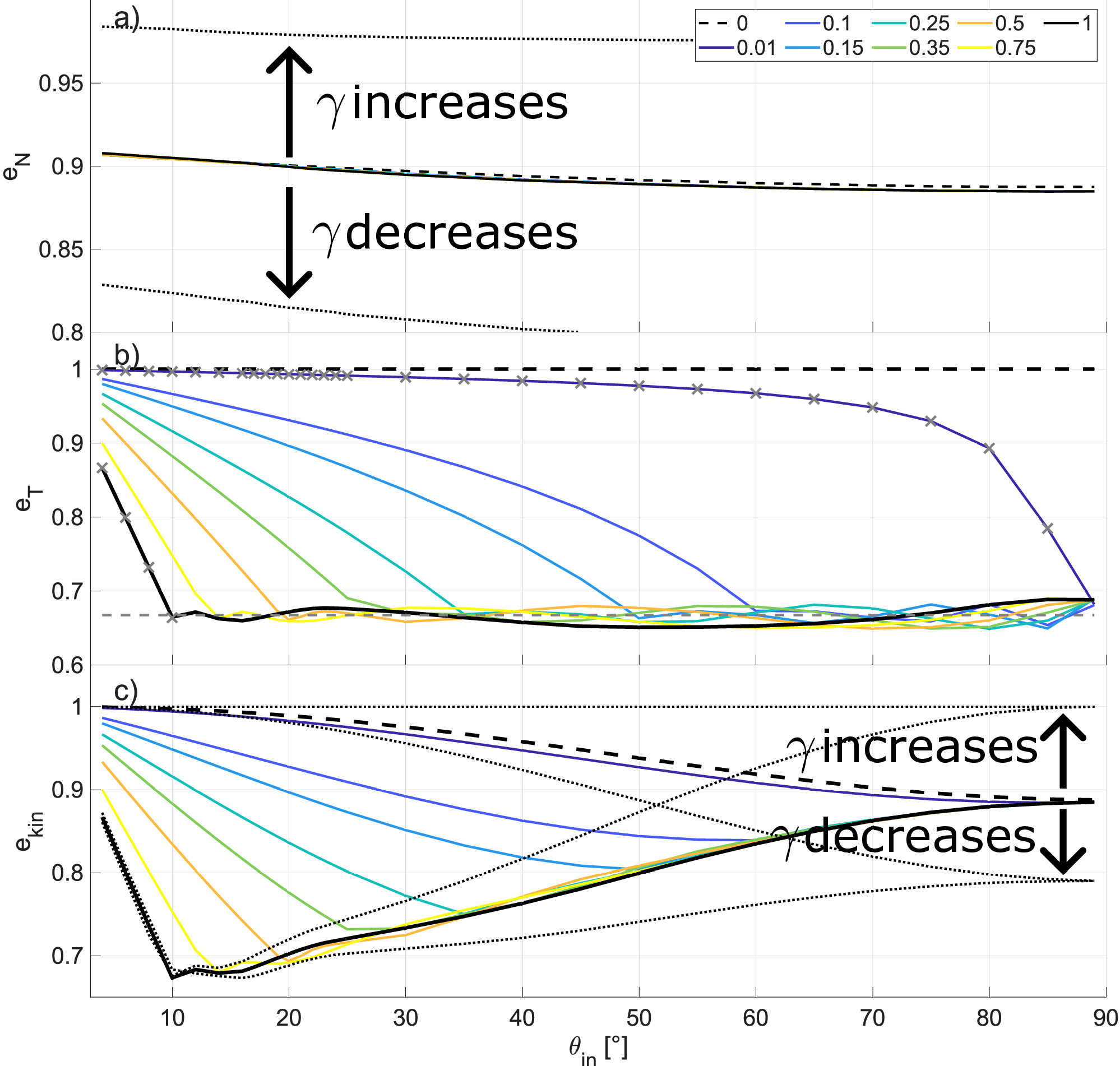}
 \caption{a) Normal $e_{\mathrm{N}}$, b) tangential $e_{\mathrm{T}}$ and c) kinematic $e_{\mathrm{kin}}$ restitution coefficients. The dotted lines represent $\mu=0$ and $\mu=1$ with different values for the normal damping $\gamma$.}
 \label{fig:02_N120_eNT}
\end{figure}

\textsl{Results.} Figure\,\ref{fig:02_N120_eNT} shows the evolution of the different restitution coefficients with the impact angle. Normal restitution $e_{\mathrm{N}}$ decreases slightly for steeper impact angles and is independent of $\mu$ as long as $\mu>0$. Increasing or decreasing the normal dissipation shifts the curve upward for lower $\gamma$ or downward for larger $\gamma$. Vanishing friction leads to a slight increase in $e_{\mathrm{N}}$ as the presence of friction can extend the contact duration, and therefore its absence limits the time for dissipation in the normal degrees of freedom.

The tangential restitution coefficient $e_{\mathrm{T}}$ shows no dependence on $\gamma$, but is instead controlled by $\mu$.  It is bounded by the case of pure sliding ($\mu=0$) and pure rolling ($\mu=\infty$). For $\theta\rightarrow0$, all $e_{\mathrm{T}}(\mu)$ converge to the same value of $e_{\mathrm{T}} = 1$, but cannot reach it since $F_{\mathrm{N}}(\theta=0^{\circ}) = 0$ and therefore $F_{\mathrm{T}}(\theta=0^{\circ}) = \mu F_{\mathrm{N}} =0$. At a critical angle $\theta\geq\theta_{\mathrm{c}}$ all  $e_{\mathrm{T}}$ collapse onto the case for pure rolling as all available translational tangential energy is converted into rotational motion. The collapse occurs earlier if $\mu$ is larger due to a faster rate of conversion. Until the beginning of the rolling regime $e_{\mathrm{T}}$ is well described by\,\cite{Sondergaard1990}
\begin{equation}
 e_{\mathrm{T}} = 1-\mu(1+e_{\mathrm{N}})\mathrm{cot}(90^{\circ}-\theta).
 \label{eq:eT_derivation}
\end{equation}
In the rolling regime this relation quickly decays to very low, and even negative values that do not represent the actual behaviour of $e_{\mathrm{T}}$. No functional relation for $e_{\mathrm{T}}$ exists for the rolling regime. Instead, $e_{\mathrm{T}}$ oscillates around a fixed value associated with pure rolling, so it cannot be described with a single value. As they can vary by approximately $2\%$ (or even larger, depending on other material properties) they are not negligible either. These oscillations occur for both perfect discs and polygons with high corner numbers, so they are also insensitive to surface roughness. Rather, surface roughness adds additional noise onto these oscillations, which shifts the values of $\theta$ at which the oscillations peak. Instead they result from a competition between rolling and sliding at the contact point, where pure rolling is interrupted by micro-slip events\,\cite{Johnson1985}. Their frequency also appears to increase with increasing angular distance to the onset of the rolling regime. If, instead of $e_{\mathrm{N}}(\theta)$, we use a constant $e_{\mathrm{N}}$, Eq.\,\ref{eq:eT_derivation} still describes the data, though the agreement with the measured values decreases the further our prescribed $e_{\mathrm{N}}$ is from the measured values.

The kinematic restitution coefficient $e_{\mathrm{kin}}$ has a more complex dependence on $\theta$. For very small $\mu$ (i.e. negligible rolling), it decreases monotonically from $e_{\mathrm{kin}}=1$ to a $\gamma$-dependent fixed value. As $\mu$ increases, $e_{\mathrm{kin}}$ decreases up to a minimum, then increases again to the $\gamma$-dependent fixed value at high $\theta$, i.e. the entire functional dependence becomes non-monotonic. As with $e_{\mathrm{T}}$, $e_{\mathrm{kin}}$ is bounded by the cases of pure sliding and pure rolling, upon which all curves for $\mu\gtrsim0.1$ collapse. $\gamma$ does not affect the initial decrease at low $\theta$, but shifts the rate of increase in the rolling regime. Like with $e_{\mathrm{T}}$, in the rolling regime, oscillations due to micro-slip  at the contact point still appear. However, unlike with $e_{\mathrm{T}}$ the oscillations in $e_{\mathrm{kin}}$ disappear quickly after $\theta_{\mathrm{c}}$ as the relative importance of the normal velocity component increases.

\begin{figure}[b]
 \centering
 \includegraphics[width=\columnwidth]{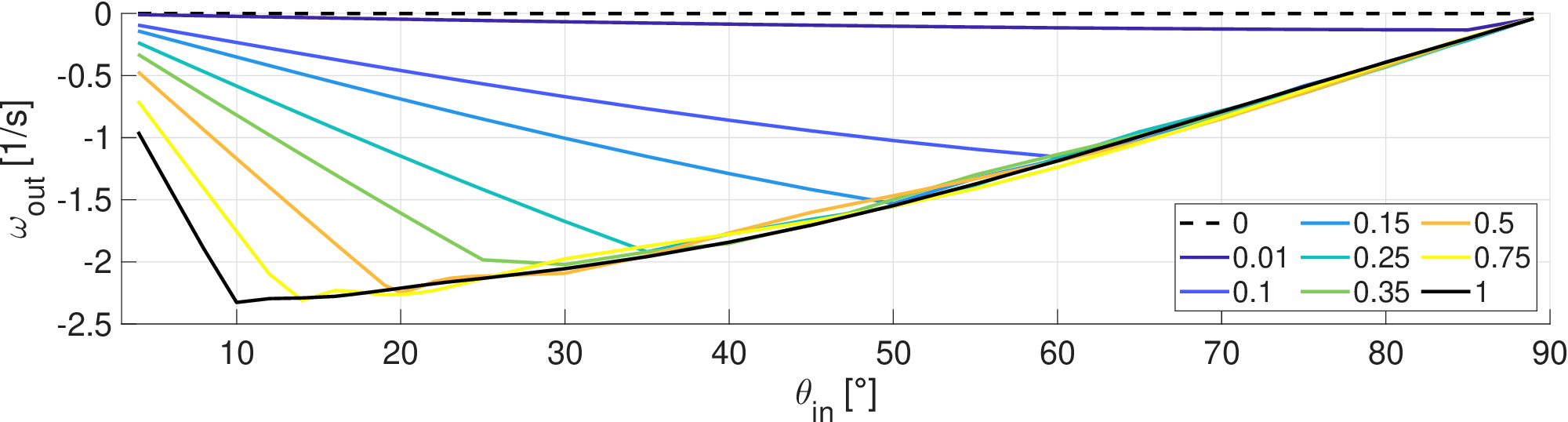}
 \caption{Angular velocity $\omega$ induced by the collision.}
 \label{fig:02_N120_v_wout}
\end{figure}
We further find a strong energy transfer from the tangential degree of freedom into rotation (Fig.\,\ref{fig:02_N120_v_wout}), which follows the overall behaviour of $e_{\mathrm{kin}}$ (agreeing with the results in\,\cite{Gorham2000}). As the impact angle $\theta$ increases, stronger rotation is induced, until a $\mu$-dependent maximum angular velocity at $\theta=\theta_{\mathrm{c}}$. For $\theta\geq\theta_{\mathrm{c}}$, frictional coupling between tangential forces and rotation eventually converts almost all available tangential velocity into rotation, so all curves collapse onto the same form. For all $\mu>0$, the induced rotation is in forward direction (clockwise, $\omega<0$) due to the tangential forces at the contact point pointing against the direction of travel.
The occurrence of maximum $\omega$ is concurrent with minimal $e_{\mathrm{kin}}$, i.e. maximal loss of energy in the translational degrees of freedom, so that $e_{\mathrm{kin}}$ is less predictive for the overall loss of energy in the system.
As with $e_{\mathrm{T}}$ and $e_{\mathrm{kin}}$ the angular velocity converges to zero at $\theta = 0^{\circ}$ and $\theta = 90^{\circ}$ due to vanishing normal and tangential forces, respectively. In contrast to $e_{\mathrm{kin}}$, $\omega$ still has a local maximum for small, but finite $\mu$.

\begin{figure}[h]
 \centering
 \includegraphics[width=\columnwidth]{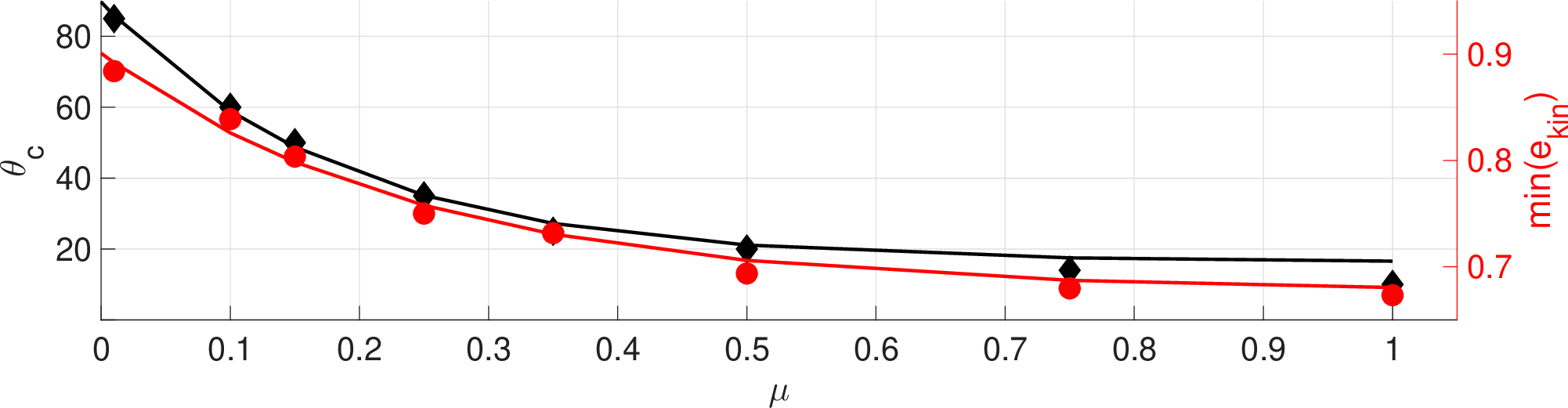}
 \caption{Value of the critical angle $\theta_c$ (left axis) and value (right axis) of the corresponding minimum restitution coefficient.}
 \label{fig:03_N120_mine}
\end{figure}

The critical angle $\theta_{\mathrm{c}}$ when the impact enters its rolling regime (i.e. maximal $|\omega|$, minimal $e_{\mathrm{kin}}$), and value of the minimum kinematic restitution coefficient, both decrease monotonically with $\mu$ (Fig.\,\ref{fig:03_N120_mine}). Their dependence can roughly be approximated with an exponential function of the form
\begin{eqnarray}
   e_{\mathrm{kin,min}} &=& a_{e}\mathrm{exp}(-b_{e}\mu)+c_{e},\\
   \theta_{\mathrm{c}}  &=& a_{\theta}\mathrm{exp}(-b_{\theta}\mu)+c_{\theta},  
\end{eqnarray}
where $a_{e} = 0.22\pm0.03$, $b_{e} = 4.07\pm1.39$, $c_{e} = 0.68\pm0.02$, and $a_{\theta} = 73.45\pm4.58$, $b_{\theta} = 5.44\pm0.79$, $c_{\theta} = 16.26\pm2.85$, although, as the variation in the fit-parameters shows, the relationship is not strictly exponential.

\begin{figure}[t]
 \centering
 \includegraphics[width=\columnwidth]{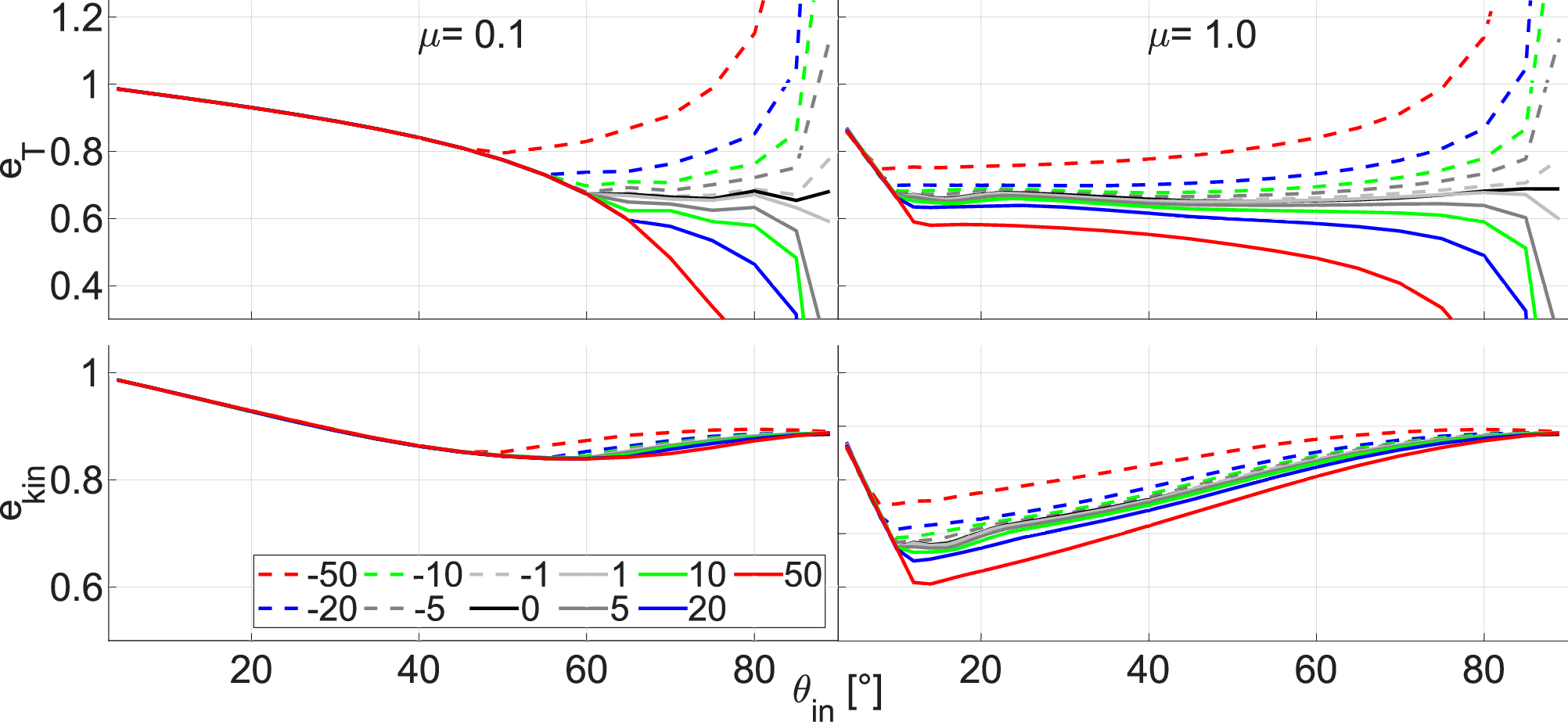}
 \caption{Influence of initial angular velocity $\omega_{\mathrm{in}} \in [-50,+50]\,\circ/s$ on $e_{\mathrm{T}}$ (above) and $e_{\mathrm{kin}}$ (below) for $\mu=0.1$ (left) and $\mu=1.0$ (right).}
 \label{fig:04_N120_eTekin_omega}
\end{figure}

While theory and simulation can and do prescribe no initial rotation of the particle, in practice (both experiment and simulation) there will almost always be a finite $\omega_{\mathrm{in}}$ upon impact. As Fig.\,\ref{fig:04_N120_eTekin_omega} shows, $\omega_{\mathrm{in}}$ can significantly alter the restitution coefficients. While the normal coefficient of restitution is unaffected by initial angular velocity beyond some small noise at low $\theta$, the situation for $e_{\mathrm{T}}$ and $e_{\mathrm{kin}}$ is different due to coupling between rotation and tangential forces. In general, clockwise rotation upon impact (CW, $\omega<0$, in direction of travel) will increase the value of the restitution coefficients as it adds to the tangential velocity of the contact point $v_{\mathrm{CP,T}}$, while counter-clockwise rotation (CCW, $\omega>0$, against direction of travel) will decrease it. Further, the magnitude and sign of $\omega_{\mathrm{in}}$ determines the critical $\theta_{\mathrm{c}}$ at which the various restitution coefficients branch off from $e(\omega_{\mathrm{in}}=0)$: at lower $\theta$ for CW rotation and at higher $\theta$ for CCW rotation, further away from $\theta_{\mathrm{c}}(\omega=0)$ for larger $|\omega_{\mathrm{in}}|$, closer for smaller $|\omega_{\mathrm{in}}|$.

For the tangential coefficient of restitution, CW rotation can lead to $e_{\mathrm{T}}\gg 1$ at steep impact angles $\theta\geq80^{\circ}$ for even very small $\omega\sim -5\, \circ$/s. Conversely, CCW rotation will lead to $e_{\mathrm{T}}\ll 0$, deflecting the particle backwards. In our case the extremal values of $e_{\mathrm{T}}$ range from $-4.17$ to $5.37$, with higher values of $\mu$ slightly reducing their magnitude. 
Rotation couples with the tangential sliding velocity at the contact point,
\begin{equation}
 v_{c,T} = v_{T}+r\omega_{\mathrm{in}},
\end{equation}
with the center-of-mass tangential velocity $v_{T}$. 
The additional contribution from rolling to $v_{c,T}$ therefore either amplifies the effect of friction by reducing the effective tangential velocity if $\omega_{\mathrm{in}}>0$, or subdues it by increasing the effective tangential velocity if $\omega_{\mathrm{in}}<0$. The larger the impact angle is, i.e. the smaller the relative value of the tangential velocity component, the stronger is this effect. Consequently, this means that the overall functional dependence of $e_{\mathrm{T}}$ on $\theta$ can become non-monotonic when considering initial particle rotation, even if discounting the micro-slip oscillations in the rolling regime for $\omega_{\mathrm{in}}=0$. Friction does not fundamentally alter this behaviour. Interestingly though, although $\theta_{\mathrm{c}}$ depends on $\mu$, the approximate value of $\theta$, from whence $e_{\mathrm{T}}$ rapidly increases seems to be independent of $\mu$, occurring at an impact angle of approximately $\theta=80^{\circ}$.
The implication of these observations is, that, while for $\omega_{\mathrm{in}}=0$, the tangential restitution coefficient could (at least in principle when omitting the micro-slip) be dealt with by a case separation according to the impact angle $\theta$, the same approach is no longer valid when the particle rotates. In other words, the tangential restitution coefficient can no longer be expressed as a function of the normal restitution, while at the same time it also varies too much to be captured by a single value. For example, in the case shown in Fig.\,\ref{fig:04_N120_eTekin_omega}, $e_{\mathrm{T}}=0.9$ corresponds to an impact angle of $\theta_{\mathrm{in}}\sim 30^{\circ}$ for all $\omega_{\mathrm{in}}$, but also 
$\theta_{\mathrm{in}}\sim 70^{\circ}$ at $\omega_{\mathrm{in}}=-50\,\circ/\mathrm{s}$, $\theta_{\mathrm{in}}\sim 83^{\circ}$ at $\omega_{\mathrm{in}}=-20\,\circ/\mathrm{s}$, $\theta_{\mathrm{in}}\sim 85^{\circ}$ at $\omega_{\mathrm{in}}=-10\,\circ/\mathrm{s}$, and $\theta_{\mathrm{in}}\sim 87^{\circ}$ at $\omega_{\mathrm{in}}=-5\,\circ/\mathrm{s}$. Even if we take a fixed angular velocity, the coefficient is not necessarily unique. Lastly, we also note that increasing $|\omega_{\mathrm{in}}|$ also minimizes the micro-slip noise in the rolling regime as the initial angular velocity dominates the competition between sliding and rolling at the contact point.

For the kinematic restitution coefficient $e_{\mathrm{kin}}$ the dependence on $\omega$ is easier: rotation still shifts the onset of the rolling regime and increases or decreases the value of $e_{\mathrm{kin}}$ therein. The overall functional form, however, remains, and for $\theta\rightarrow90^{\circ}$, all $e_{\mathrm{kin}}(\omega_{\mathrm{in}})$ converge onto the same value due to the vanishing portion of tangential forces in the interaction. Increasing friction does not change this behaviour, but only increases by how much $|\omega_{\mathrm{in}}|$ changes $e_{\mathrm{kin}}$. Regardless of $|\omega_{\mathrm{in}}|$, $e_{\mathrm{kin}}$ remains non-monotonic, with a local minimum at a value of $\theta$ that now depends on both $\mu$ and $\omega$.

\textsl{Implications for numerical simulations}
Many high-performance simulations compute their interactions by prescribing a single fixed value for the different restitution coefficients as dissipative material parameters, or at least derive their material parameters in some form from the restitution coefficients. 

For frictionless hard-sphere systems this may be acceptable: Interaction occurs only in normal direction and the angular degrees of freedom are decoupled from the rectilinear motion. The velocity dependence of $e_{\mathrm{N}}$ can be taken into account, and the contact direction dependence is small enough to probably be negligible.
However, real materials are always frictional and so interactions in other directions must be accounted for. Taking a fixed value obtained from some experiment only describes a non-unique situation, not a material. As shown here, the same value can be mapped onto a wide variety of $(\theta, \omega_{\mathrm{in}}, \mu)$-combinations and so is highly situational at best. Even for the same impact conditions, varying the impact direction alone can lead to a significant change in tangential restitution (e.g. changing by $\pm 4$ times in our simulations).
Deriving a unique $e_{\mathrm{T}}$ from a unique $e_{\mathrm{N}}$ is not a viable approach either, as Eq.\,(\ref{eq:eT_derivation}) is valid only up to a limit, after which it utterly fails. Further, it does not even take into account the initial particle rotation which leads to even broader distributions of the restitution coefficients. Moreover, it cannot represent the (physically correct\,\cite{Johnson1985}) micro-slip at steeper impact angles.
If the restitution coefficients used as input parameters do not map uniquely to a material, but rather are highly situational parameters, then questions arise about the physical validity of simulations using these parameters and the resulting predictions.

We can go further: If the use of restitution coefficients is already problematic for round particles, how do these coefficients perform for non-round particles. Most granular particles are non-spherical, so accounting for shape is a must. However, taking particle shape into account will only compound the complications, as the particle orientation upon impact also couples with the impact kinematics\,\cite{Krengel2026}, even without initial rotation, an issue that only worsens with the additional degrees of freedom in three dimensions.

\textsl{Conclusions}
We have performed a detailed study of the various coefficients of restitution of spherical particles in a planar collision by means of discrete element simulations across a wide range of initial conditions. We found that, while the normal coefficient of restitution is essentially unique for each material, the tangential and kinematic coefficients of restitution are not unique material parameters, but rather highly situational parameters with non-unique material dependence. Our results indicate that restitution coefficients are ill-suited as control parameters for numerical simulations and, besides normal restitution, poorly suited for the description of granular systems. The impact of a disc is simple in two dimensions, the axis of rotation is always perpendicular to the plane spanned by the normal and tangential velocity components, and so rotation always maps onto the direction of travel. The same is not true in three dimensions: While the translational motion can still be reduced to a two-dimensional normal and tangential direction pair, rotation can occur at any angle to the direction of travel, which complicates extending the information presented here to three dimensions. Further, in three dimensions it is possible for the particle to pivot while sliding\,\cite{Shegelski2018}, which further alters the particle kinematics and has a much higher sensitivity to the interface conditions than rolling. How the restitution coefficients evolve with the impact and sliding conditions in three dimensions remains to be studied. During the collision between two discs, the mutual center-of-mass may rotate which distorts the rebound kinematic variables in a global frame of reference. Nevertheless, our results should still hold true to some extent in a local frame of reference. 
In this work we have measured only the kinematic variables before and after the impact, not how they achieve their specific values, i.e. the details of the contact process are irrelevant to the analysis performed here. This means our results are independent of the particular contact model one may choose for their simulations, and apply universally to planar impacts of discs.

\begin{acknowledgments}
\textsl{Acknowledgements}
DK is grateful for the support by Professor Shun Nomura of Tokyo University of Marine Science and Technology and to Professor Hans-Georg Matuttis for many helpful comments during the writing of this manuscript.
\end{acknowledgments}

\bibliography{impact_rotation_disc}
\end{document}